# PERSIAN VERSION OF WAYFINDING QUESTIONNAIRE

**Mobina Zibandehpoor**
Department of Mechatronics
Faculty of Electrical Engineering
K. N. Toosi University of Technology
Tehran, Iran
m.zibandepour@email.kntu.ac.ir

**Mehdi Delrobaei**
Department of Mechatronics
Faculty of Electrical Engineering
K. N. Toosi University of Technology
Tehran, Iran
delrobaei@kntu.ac.ir

## ABSTRACT

Spatial navigation ability is essential for daily functioning, and the Wayfinding Questionnaire (WQ) is a validated self-report tool assessing this ability through 22 items across three subscales: Navigation and Orientation (11 items), Distance Estimation (3 items), and Spatial Anxiety (8 items). This study introduces the Persian translation of the WQ, adapted for Persian-speaking populations using a rigorous "forward-backward" translation, cognitive debriefing, and cultural adaptation process to ensure alignment with the original tool's reliability and validity. The Persian WQ provides a complete assessment of spatial navigation skills and identifies potential navigation challenges. Furthermore, the Persian WQ serves as a valuable resource for future research exploring spatial navigation and memory in diverse populations.

## 1 Introduction

Navigation ability, the capacity to orient oneself and move purposefully within the environment, is fundamental to human autonomy and daily functioning, enabling individuals to interact confidently and effectively with their surroundings. This cognitive skill encompasses a range of mental processes, including spatial memory—recalling locations and landmarks; path planning—formulating efficient routes; and spatial awareness—understanding distances, directions, and relational positioning within a given space Muffato et al. [2022]. These interconnected abilities allow individuals to navigate familiar routes and adapt to new and complex environments. Impairments in navigation ability can significantly restrict independence, posing challenges in essential activities such as driving, traveling, or managing unfamiliar locations, thereby reducing the overall quality of life Halko et al. [2014]. Deficits in these areas can lead to increased dependency on others, heightened anxiety, and a diminished sense of autonomy.

Self-report questionnaires are widely used to assess navigation abilities, offering an efficient, standardized way to capture subjective navigation challenges that may not be fully revealed through objective tests. These tools are particularly valuable for understanding individual experiences with spatial orientation, distance estimation, and related anxieties, which contribute to overall navigation ability Hegarty et al. [2002]. The Wayfinding Questionnaire (WQ) is one such validated tool, consisting of 22 items across three subscales: Navigation and Orientation, Distance Estimation, and Spatial Anxiety. Developed to identify navigation complaints, the WQ provides insight into the cognitive and emotional components of navigation ability Claessen et al. [2016]. The WQ has been adapted into several languages to facilitate cross-cultural research on spatial abilities. Originally developed by the van der Ham research group in the Netherlands, a validated Dutch version exists, and a recent study by a Turkish research group has adapted and validated the WQ for Turkish-speaking populations. These efforts highlight the questionnaire's utility across different languages and cultural contexts, underscoring its applicability in diverse populations Avcı and Aksoy [2024].

By translating and culturally adapting the WQ into Persian, this study aims to broaden access to a validated tool for assessing navigation abilities, enabling Persian-speaking individuals to self-assess spatial navigation skills and related challenges. This adaptation provides researchers and clinicians with a reliable means of evaluating navigation abilities within Persian-speaking populations, fostering an improved understanding of navigation-related complaints and supporting targeted interventions. In addressing a gap in culturally sensitive assessment tools, our study lays the groundwork for future cross-cultural research on navigation abilities and opens doors for additional language adaptations.

## 2  Method

In the following section, we provide a comprehensive overview of the translation procedure, the scaling validation process, and the detailed scoring methodology for each subscale, essential steps to ensure both the linguistic integrity and measurement reliability of the Persian version of the WQ (Appendix1).

### 2.1  Translation Procedure

The translation of the WQ adhered to a rigorous "forward-backward" methodology to achieve both linguistic and conceptual fidelity. In the initial phase, two bilingual experts independently translated the 22 items from English to Persian, creating a preliminary Persian version. This version underwent a comprehensive review by a third expert, who reconciled any inconsistencies and refined the language to improve clarity and cultural relevance. Following these refinements, an independent bilingual translator back-translated the Persian version into English. This back-translation process allowed for a detailed comparison with the original English version, facilitating the identification of any nuanced discrepancies in meaning or intent. To further validate the Persian version's comprehensibility and ensure alignment with the original content, three Persian-speaking participants reviewed and completed the final questionnaire. This final review confirmed that the translation maintained conceptual integrity and was culturally accessible for Persian-speaking users, ensuring both linguistic accuracy and practical applicability.

### 2.2  Scoring

The WQ was evaluated using a Likert-scale approach, aimed at assessing the extent of reliance on different navigation strategies. Each item on the original WQ uses a 7-point scale, ranging from "Not at all applicable to me" to "Fully applicable to me." This scale allows for the summing of scores across items, generating composite scores for each subscale and offering a detailed evaluation of personal navigation patterns. Items 12, 13, and 14 specifically use a scale where a score of 1 represents "not uncomfortable at all" and a score of 7 represents "very uncomfortable". To account for spatial anxiety in particular, reverse scoring was applied to relevant items. This means that for items associated with anxiety about spatial navigation, lower scores indicate higher levels of navigation-related complaints. This reverse scoring adjustment provides a more accurate measure of spatial anxiety by aligning lower scores with increased anxiety levels.

## 3  Conclusion

In conclusion, this study successfully adapted the WQ for Persian-speaking populations, ensuring its reliability and validity through a thorough translation and cultural adaptation process. The Persian WQ offers a robust tool for assessing spatial navigation skills and identifying challenges within this linguistic group, promoting more accurate self-assessment and supporting tailored interventions. The Persian WQ is also a valuable resource for researchers and clinicians investigating spatial cognition in diverse populations. In future work, statistical analyses can further evaluate the questionnaire's psychometric properties, enhancing its utility and providing cross-cultural insights into spatial navigation abilities.

**Appendix 1**

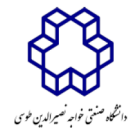

<div dir="rtl">

# پرسشنامه مسیریابی

۲۲ عبارت زیر در مورد توانایی ناوبری مکانی شما در زندگی روزمره است. لطفاً برای هر یک از این موارد، عددی را که به بهترین وجه توانایی شمارا توصیف می‌کند، علامت بزنید.

| ۷ | ۶ | ۵ | ۴ | ۳ | ۲ | ۱ |
|---|---|---|---|---|---|---|
| کاملاً در مورد من صدق می‌کند. | تقریباً همیشه در مورد من صدق می‌کند. | غالباً در مورد من صدق می‌کند. | گاهی در مورد من صدق می‌کند. | به ندرت در مورد من صدق می‌کند. | تقریباً هرگز در مورد من صدق نمی‌کند. | اصلاً در مورد من صدق نمی‌کند. |

۱. وقتی برای اولین بار در یک ساختمان هستم، به‌راحتی می‌توانم به ورودی اصلی این ساختمان اشاره‌کنم.

کاملاً در مورد من صدق می‌کند. ⑦ ⑥ ⑤ ④ ③ ② ① اصلاً در مورد من صدق نمی‌کند.

۲. اگر چندین بار یک مکان (ساختمان، بنای تاریخی، تقاطع) را ببینم، دقیقاً می‌دانم که قبلاً از کدام سمت آن را دیده بودم.

کاملاً در مورد من صدق می‌کند. ⑦ ⑥ ⑤ ④ ③ ② ① اصلاً در مورد من صدق نمی‌کند.

۳. در شهری ناآشنا وقتی نقشه‌ای را از تابلوی اطلاعات می‌خوانم به‌راحتی می‌توانم بفهمم کجا باید بروم.

کاملاً در مورد من صدق می‌کند. ⑦ ⑥ ⑤ ④ ③ ② ① اصلاً در مورد من صدق نمی‌کند.

۴. بدون نقشه می‌توانم مسافت مسیری را که برای اولین بار پیموده‌ام به‌خوبی تخمین بزنم.

کاملاً در مورد من صدق می‌کند. ⑦ ⑥ ⑤ ④ ③ ② ① اصلاً در مورد من صدق نمی‌کند.

۵. زمانی که نقشه یک شهر ناآشنا را باراهنما و مقیاس ببینم، می‌توانم به‌خوبی تخمین بزنم که چقدر طول می‌کشد تا مسیر پیاده‌روی را طی کنم.

کاملاً در مورد من صدق می‌کند. ⑦ ⑥ ⑤ ④ ③ ② ① اصلاً در مورد من صدق نمی‌کند.

۶. وقتی در محیطی ناآشنا هستم همیشه می‌توانم به‌سرعت و به‌درستی جهت‌یابی کنم.

کاملاً در مورد من صدق می‌کند. ⑦ ⑥ ⑤ ④ ③ ② ① اصلاً در مورد من صدق نمی‌کند.

۷. من همیشه می‌خواهم بدانم دقیقاً کجا هستم (همیشه سعی می‌کنم در محیط‌های ناآشنا خودم را پیدا کنم.)

کاملاً در مورد من صدق می‌کند. ⑦ ⑥ ⑤ ④ ③ ② ① اصلاً در مورد من صدق نمی‌کند.

۸. من از اینکه درجایی راه خود را گم کنم، نگران می‌شوم.

کاملاً در مورد من صدق می‌کند. ⑦ ⑥ ⑤ ④ ③ ② ① اصلاً در مورد من صدق نمی‌کند.

۹. من ازینکه در شهری ناآشنا گم شوم دچار نگرانی می‌شوم.

کاملاً در مورد من صدق می‌کند. ⑦ ⑥ ⑤ ④ ③ ② ① اصلاً در مورد من صدق نمی‌کند.

</div>

۱۰. در یک شهر ناآشنا، ترجیح می‌دهم به‌جای تنهایی قدم زدن، با گروه قدم بزنم.

| اصلاً در مورد من صدق نمی‌کند. | ⑦ | ⑥ | ⑤ | ④ | ③ | ② | ① | کاملاً در مورد من صدق می‌کند. |
|---|---|---|---|---|---|---|---|---|

۱۱. گم‌شدن باعث اضطراب من می‌شود.

| اصلاً در مورد من صدق نمی‌کند. | ⑦ | ⑥ | ⑤ | ④ | ③ | ② | ① | کاملاً در مورد من صدق می‌کند. |
|---|---|---|---|---|---|---|---|---|

❖ در موقعیت‌های زیر (موارد ۱۲، ۱۳ و ۱۴) چقدر احساس اضطراب می‌کنید؟

۱۲. تصمیم‌گرفتن برای تعین مسیر پس از پیاده شدن از وسیله حمل‌ونقل عمومی مانند قطار اتوبوس مترو

| اصلاً مضطرب نمی‌شوم. | ⑦ | ⑥ | ⑤ | ④ | ③ | ② | ① | بسیار مضطرب می‌شوم. |
|---|---|---|---|---|---|---|---|---|

۱۳. پیدا کردن مسیر در یک ساختمان ناآشنا (مثلاً یک بیمارستان).

| اصلاً مضطرب نمی‌شوم. | ⑦ | ⑥ | ⑤ | ④ | ③ | ② | ① | بسیار مضطرب می‌شوم. |
|---|---|---|---|---|---|---|---|---|

۱۴. پیدا کردن مسیر برای رفتن به یک جلسه در یک شهر ناآشنا.

| اصلاً مضطرب نمی‌شوم. | ⑦ | ⑥ | ⑤ | ④ | ③ | ② | ① | بسیار مضطرب می‌شوم. |
|---|---|---|---|---|---|---|---|---|

۱۵. رفتن به مقصدی که قبلاً نرفته‌ام برایم نگران‌کننده است.

| اصلاً در مورد من صدق نمی‌کند. | ⑦ | ⑥ | ⑤ | ④ | ③ | ② | ① | کاملاً در مورد من صدق می‌کند. |
|---|---|---|---|---|---|---|---|---|

۱۶. من معمولاً می‌توانم یک مسیر جدید را بعد از یک‌بار پیاده‌روی به یاد بیاورم.

| اصلاً در مورد من صدق نمی‌کند. | ⑦ | ⑥ | ⑤ | ④ | ③ | ② | ① | کاملاً در مورد من صدق می‌کند. |
|---|---|---|---|---|---|---|---|---|

۱۷. من از توانایی مناسبی در تخمین فاصله ها برخوردارم (مثلاً از جایی که هستم تا ساختمانی که می توانم ببینم)

| اصلاً در مورد من صدق نمی‌کند. | ⑦ | ⑥ | ⑤ | ④ | ③ | ② | ① | کاملاً در مورد من صدق می‌کند. |
|---|---|---|---|---|---|---|---|---|

۱۸. من در فهم شرح مسیر و دنبال کردن آن خوب عمل می‌کنم.

| اصلاً در مورد من صدق نمی‌کند. | ⑦ | ⑥ | ⑤ | ④ | ③ | ② | ① | کاملاً در مورد من صدق می‌کند. |
|---|---|---|---|---|---|---|---|---|

۱۹. من در توضیح دادن مسیرها (یعنی بیان مسیرهای شناخته شده برای شخص دیگر) مهارت دارم.

| اصلاً در مورد من صدق نمی‌کند. | ⑦ | ⑥ | ⑤ | ④ | ③ | ② | ① | کاملاً در مورد من صدق می‌کند. |
|---|---|---|---|---|---|---|---|---|

۲۰. وقتی از فروشگاهی خارج می شوم، نیازی نیست دوباره جهت یابی کنم تا بفهمم کجا باید بروم.

| اصلاً در مورد من صدق نمی‌کند. | ⑦ | ⑥ | ⑤ | ④ | ③ | ② | ① | کاملاً در مورد من صدق می‌کند. |
|---|---|---|---|---|---|---|---|---|

۲۱. من از انتخاب مسیرهای جدید (به عنوان مثال راه های میانبر) به مقصدهای آشنا لذت می برم.

| اصلاً در مورد من صدق نمی‌کند. | ⑦ | ⑥ | ⑤ | ④ | ③ | ② | ① | کاملاً در مورد من صدق می‌کند. |
|---|---|---|---|---|---|---|---|---|

۲۲. من به راحتی می توانم کوتاه ترین مسیر را برای رسیدن به یک مقصد آشنا پیدا کنم.

| اصلاً در مورد من صدق نمی‌کند. | ⑦ | ⑥ | ⑤ | ④ | ③ | ② | ① | کاملاً در مورد من صدق می‌کند. |
|---|---|---|---|---|---|---|---|---|